\documentclass[MSNbibl,nameyear,dvips]{arxstspdf}
\usepackage{graphicx}
\usepackage{flushend}
\usepackage{stfloats}

\volume{28}
\issue{4}
\pubyear{2013}
\firstpage{521}
\lastpage{541}
\doi{10.1214/13-STS439} 

\makeatletter
\renewcommand{\textdegree}{${}^\circ$}
\makeatother

\begin{document}
\begin{frontmatter}

\title{A Prospect of Earthquake Prediction
Research}%
\runtitle{Earthquake Prediction}

\begin{aug}
\author[a]{\fnms{Yosihiko} \snm{Ogata}\corref{}\ead[label=e1]{ogata@ism.ac.jp}}
\runauthor{Y. Ogata}

\affiliation{Information and System Research Organization}

\address[a]{Yosihiko Ogata is Professor Emeritus,
Institute of Statistical Mathematics,
Information and System Research Organization,
10-3 Midori-cho, Tachikawa, Tokyo 190-8562
and
Institute of Industrial Science,
University of Tokyo, Visiting Professor,
4-6-1 Komaba, Megro-Ku, Tokyo 153-8505 \printead{e1}.}
\end{aug}

\begin{abstract}
Earthquakes occur because of abrupt slips on faults due to accumulated
stress in the Earth's crust. Because most of these faults and their
mechanisms are not readily apparent, deterministic earthquake prediction is
difficult. For effective prediction, complex conditions and uncertain
elements must be considered, which necessitates stochastic prediction. In
particular, a~large amount of uncertainty lies in identifying whether
abnormal phenomena are precursors to large earthquakes, as well as in
assigning urgency to the earthquake. Any discovery of potentially useful
information for earthquake prediction is incomplete unless quantitative
modeling of risk is considered. Therefore, this manuscript describes the
prospect of earthquake predictability research to realize practical
operational forecasting in the near future.
\end{abstract}


\end{frontmatter}

\section{Introduction}\label{sec1}

Through remarkable developments in solid Earth science since the late
1960s, our understanding of earthquakes has increased significantly. The
availability of relevant data has steadily increased as the study of
earthquakes has progressed remarkably in geophysics. After every major
earthquake, researchers have elucidated important seismic mechanisms
associated with it. However, even though detailed analysis and discussions
have been conducted, large uncertainties remain because of diversity and
complexity of the earthquake phenomenon. This leads to unachievable
challenges in deterministic earthquake prediction because all diverse and
complex scenarios must faithfully reflect the processes of earthquakes to
be considered for effective earthquake prediction.

On the other hand, several techniques for predicting earthquakes have been
proposed on the basis of anomalies of various types; however, the
effectiveness of these techniques is controversial (\cite{JORetal11}).
Therefore, objectivity is required for such evaluation; otherwise,
arguments presented may lack merit. New prediction models that claim to
incorporate potentially useful information over those used in standard
seismicity models should be evaluated to determine whether predictive power
is improved. Earthquake forecasting models should evolve in this manner.

Recently, there has been growing momentum for seismologists to develop an
organized research program on earthquake predictability. An international
cooperative study known as \textit{Collaboratory for the Study of
Earthquake Predictability} (CSEP; \url{http://www.cseptesting.org/})
is currently under way among countries prone to major earthquakes for
exploring possibilities in earthquake prediction (e.g., \cite*{JOR06}). An
immediate objective of the project is to encourage the development of
statistical models of seismicity, such as those subsequently discussed in
Section~\ref{sec2}, and to evaluate their predictive performances in terms of
probability.

In addition, the CSEP study aims to develop a scientific infrastructure to
evaluate statistical significance and \textit{probability} \textit{gain}
(\cite{AKI81}) of various methods used to predict large earthquakes by using
observed abnormalities such as seismicity anomaly, transient crustal
movements and electromagnetic anomaly. Here \textit{probability gain} is
defined as the ratio of probability of a large earthquake estimated based
on an anomaly to the underlying probability without anomaly. Section~\ref{sec3}
describes this important concept, and then discusses statistical
point-process models to examine the significance of causality of anomalies
and also to evaluate the probability gains conditional on the anomalous
events.

For prediction of large earthquakes with a higher probability gain,
comprehensive studies of anomalous phenomena and observations of earthquake
mechanisms are essential. Several such studies are summarized in Sections~\ref{sec4}--\ref{sec6}. Particularly, I~have been interested in elucidating abnormal seismic
activities and geodetic anomalies to apply them for promoting forecasting
abilities, as described in these sections.

\section{Probability Forecasting of Baseline Seismicity}\label{sec2}

\subsection{Log-Likelihood for the Evaluation Score of Probability
Forecast}\label{sec2.1}

Through repeated revisions, CSEP attempts to establish standard models to
predict probability that conform to various parts of the world. Here I mean
the prediction/estimation of probability as predicting/estimating
conditional probabilities given the past history of earthquakes and other
possible precursors. The likelihood is used as a reasonable measure of
prediction performance (cf. \cite*{BOL78}; \cite*{Aka85}). The evaluation
method for probabilistic forecasts of earthquakes by the log-likelihood
function has been proposed, discussed and implemented (e.g., \cite*{KAGJAC95}; \cite*{O95};
Ogata, Utsu and Katsura, \citeyear{OGAUTSKAT96}; \cite*{V99}; \cite*{HARVER05};
\cite*{Setal07}; Zechar, Gerstenberger and
  Rhoades, \citeyear{ZECGERRHO10}; \cite*{NANetal12}; \cite*{OGAetal13}). In some\vadjust{\goodbreak} such studies, the evaluation score
has been referred to as (relative) entropy, which is essentially similar to
the log-likelihood.

\subsection{Space--Time-Magnitude Forecasting of
Earthquakes}\label{sec2.2}

Baseline models should be set to compare with and evaluate all
predictability models. Based on empirical laws, we can predict standard
reference probability of earthquakes in a space--time-magnitude range on
the basis of the time series of present and past earthquakes. The framework
of CSEP, which has evaluated performances of submitted forecasts of
respective regions (\cite*{JOR06}; Zechar, Gerstenberger and
  Rhoades, \citeyear{ZECGERRHO10}; \cite*{NANetal11}),
is similar to that of the California Regional Earthquake Likelihood Models (RELM)
project for spatial forecast (\cite*{F07}; \cite*{SCHetal10}). Different space--time models
were submitted to the CSEP Japan Testing Center at the Earthquake Research
Institute, University of Tokyo, for the one-day forecast applied to the
testing region in Japan (\cite{NANetal12}). This means that the model
forecasts the probability of an earthquake at each space--time-magnitude
bin. However, the CSEP protocols have to be improved to those in terms of
point-processes on a continuous time axis for the evaluation including a
real-time forecast (\cite{OGAetal13}).

Almost all models incorporated the Gutenberg--Richter (G--R) law for
forecasting the magnitude factor, and take different variants of the
space--time epidemic-type aftershock sequence (ETAS) model (\cite*{NANetal12},
and \cite*{OGAetal13}). In the following sections, I will outline
these models.

\begin{figure*}

\includegraphics{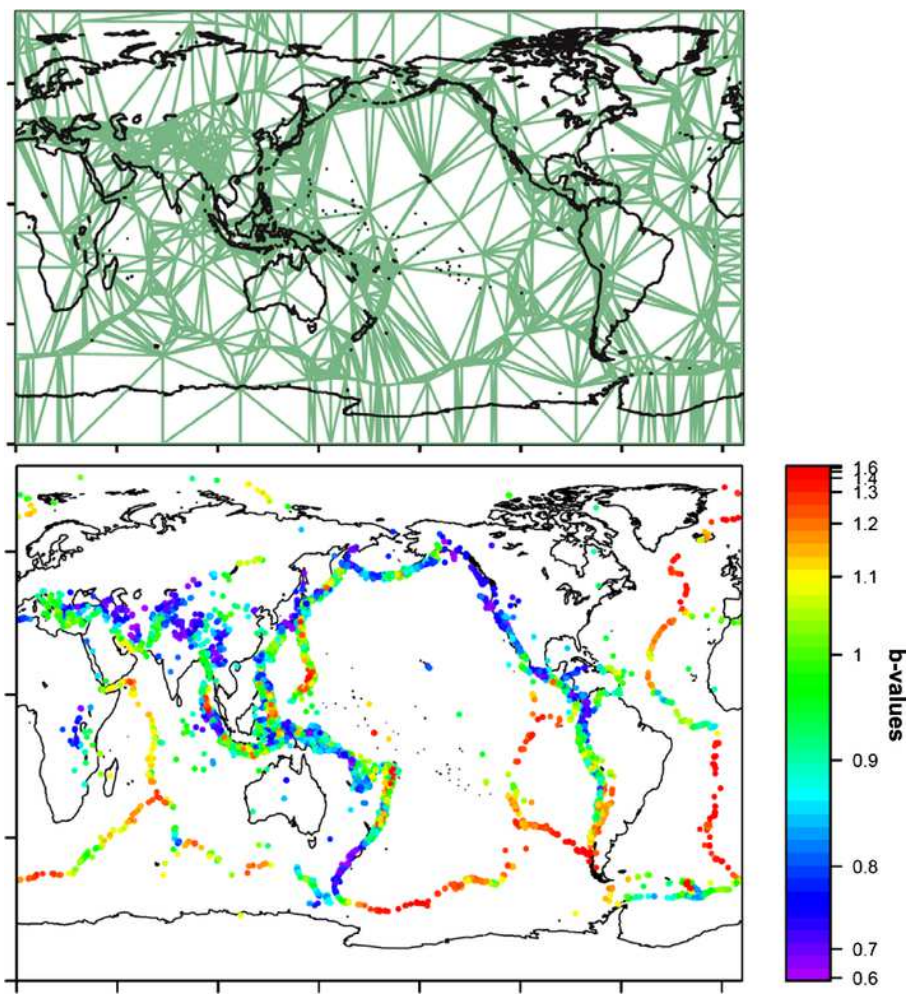}

\caption{Top panel shows Delaunay tessellation upon which the
piecewise linear function is defined. The smoothness constraint is posed in
the sum of squares of integrated slopes. Bottom panel shows b-value estimates of the
G--R formula in equation (\protect\ref{eq1}) estimated from the data for earthquakes of
$\mathrm{M}\ge5.4$ from the Harvard University global CMT
catalog (\protect\url{http://www.globalcmt.org/CMTsearch.html}). One conspicuous feature
is that b-values are large in oceanic ridge zones, but small in plate subduction
zones.}\label{f01}
\end{figure*}

\subsubsection{Magnitude frequency distribution}\label{sec2.2.1}

Gutenberg and Richter (\citeyear{GUTRIC44}) determined that the number of earthquakes
increased (decreased) exponentially as their magnitude decreased
(increased). Describing this theory in terms of point processes, the
intensity of magnitude $M$ is
%
\begin{eqnarray}\label{eq1}\quad
\lambda_0(M)&=&\lim_{\Delta\to0}\frac{1}{\Delta}\operatorname{Prob}(M<\mathit{Magnitude}\leq M+\Delta)
\nonumber
\\[-8pt]
\\[-8pt]
\nonumber
&=&10^{a-bM}=Ae^{-\beta M}
\end{eqnarray}
%
for constants $a$ and $b$. In other words, the magnitude of each earthquake
will obey an exponential distribution such that $f(M) = \beta
\operatorname{e}^{ - \beta (M - Mc)}, M \ge M_{c}$, where $\beta = b\ln 10$,
and $M_{c}$ is a cutoff magnitude value above which all earthquakes are
detected. Traditionally, the $b$-value had been estimated\vadjust{\goodbreak} graphically,
however, more efficient estimation is performed by the maximum likelihood
method. \citet{UTS65} derived it by the moment method. Later, \citet{AKI65}
demonstrated that this is a maximum-likelihood estimate (MLE) and provided
the error estimate. It should be noted that the magnitudes in most catalogs
are given in the interval of 0.1 (discrete magnitude values), hence, care
should be taken for avoiding the bias of the $b$ estimate in
likelihood-based estimation procedures (\cite{UTSN1}).

Although the coefficient $b$ in a wide region is generally slightly smaller
than 1.0, \citet{GUTRIC44} further determined that the
$b$-value varies according to location in smaller seismic regions. The
\mbox{$b$-value} varies even within Japan and further varies with time. Temporal
and spatial \mbox{$b$-value} changes have attracted the attention of many
researchers ever since \citet{SUY66} reported a difference between
$b$-values of foreshocks and aftershocks in a sequence.

Here, we consider that $\beta$ can vary with time, space and space--time
according to a function such as $\beta (t)$, $\beta (x, y, z)$ or $\beta
(t, x, y)$. Various nonparametric smoothing algorithms such as kernel
methods have been proposed (Wiemer and Wyss, \citeyear{WIEWYS97}). Alternatively, the
$\beta$ value can be parameterized by smooth cubic splines (Ogata, Imoto and Katsura,
\citeyear{OGAIMOKAT91}; \cite*{OGAKAT93}) or piece-wise linear expansions on Delaunay
tessellated space\ (\cite*{OGA11N2}; see also Figure~\ref{f01}, e.g.). In such a
case, a penalized log-likelihood (\cite*{GG71}) is used whereby the log-likelihood
function is associated with penalty functions in which the coefficients are
constrained for smoothness of the $\beta$ function. For the optimal
estimation of the $\beta$ function, the weights in the penalty function are
objectively adjusted in a Bayesian framework, as suggested by \citet{Aka80N1}.

\subsubsection{Aftershock analysis and probabilistic
forecasting}\label{sec2.2.2}

Typical aftershock frequency decays according to the reverse power function
with time (\cite*{OMO}; Utsu, \citeyear{UTS61}, \citeyear{UTSN1}). First, let $N(s,t)$ be the number
of aftershocks in an interval ($s, t$). Then, the occurrence rate of
aftershocks at the elapsed time $t$ since the main\ shock is
%
\begin{eqnarray}\label{eq2}
 \nu (t) &=& \lim_{\Delta \to 0}\frac{1}{\Delta} P \bigl\{ N(t,t + d
\Delta ) \ge 1|\nonumber\\
&&\hspace*{48pt}{}\mbox{Mainshock at time } 0 \bigr\}
\\
& =& \frac{K}{(t + c)^{p}}\nonumber
\end{eqnarray}
for constants $K, c$ and $p$. This is known as the Omori--Utsu (O--U) law.

Traditionally, estimates of the parameter $p$ have been obtained since the
study of \citet{UTS61} in the following manner. The numbers of aftershocks in
a unit time interval $n(t)$ are first plotted against elapsed time on
doubly logarithmic axes, and then are fit to an asymptotic straight line.
The slope of this line is an estimate for $p$. The values of $c$ can be
determined by measuring the bending curve immediately after the main shock.
Such an analysis is based on the time series of counted numbers of
aftershocks. By such a plot, we can find aftershock sequences for which the
formula (\ref{eq2}) lasts a long period, more than 120 years, for example (\cite*{UTSN1}; \cite*{OGA89}; Utsu, Ogata and Matsu'ura, \citeyear{UTSOGAMAT95}).

To efficiently estimate the three parameters directly on the basis of
occurrence time records of aftershocks, assuming nonstationary Poisson
process with intensity function (\ref{eq2}), \citet{OGA83} suggested the
maximum-likelihood method, which enabled the practical aftershock
forecasting. \citet{REAJON89} proposed a procedure based on the
joint intensity rate of time and magnitude of aftershocks (\cite*{U70})
according to the G--R law~(\ref{eq1}),
%
\begin{eqnarray}\label{eq3}
\quad\lambda (t,M) &= &\lambda_{0}(M)\nu (t)
\nonumber
\\[-8pt]
\\[-8pt]
\nonumber
&=& \frac{10^{a + b(M_{0} - M)}}{(t
+ c)^{p}}\quad (a, b,
c, p;\ \mathrm{constant}),
\end{eqnarray}
where $M$ is the magnitude of an aftershock and $t$ is the time following a
main shock of magnitude $M_{0}$; the parameters are independently estimated
by the maximum-likelihood method for respective empirical laws.

After a large earthquake occurs, the Japan Meteorological Agency (JMA) and
the United States Geological Survey (USGS) have undertaken operational
probability forecast of the aftershocks. However, the forecast is announced
after the elapse of 24~h or more. This is due to the deficiency of
aftershock data due to overlapping of seismograms after the main shock. In
particular, the parameter $a$ is crucial for the early forecast, but
difficult to estimate in an early period, whereas the other parameters can
be default values for the early forecast [\cite*{REAJon94};
Earthquake Research Committee (ERC), \citeyear{EAR98}]. The difficulty is because the
parameter $a$ can substantially differ even if the magnitudes of the main
shocks are almost the same: for example, the numbers of the aftershocks of
$\mathrm{M} \geq 4.0$ of two nearby main shocks of the same M6.8 differ by
6--7 times (\cite*{JMA09}).

It is notable that the strongest aftershocks occurred within 24~h in most
sequences (\cite*{JMA08}). Therefore, despite adverse conditions during data
collection, probabilistic aftershock forecasts should be delivered as soon
as possible within 24~h after the main shock to mitigate secondary disasters
in affected areas.

For this purpose, it is necessary to estimate time-dependent missing rates,
or detection rates, of aftershocks (Ogata and Katsura, \citeyear{OGAKAT93}, \citeyear{OGAKAT06}; Ogata,
\citeyear{OGA05c}) because they enable probabilistic forecasting immediately after the
main shock (Ogata, \citeyear{OGA05c}; \cite*{OGAKAT06}). The detection rate of
earthquakes is described by a probability function $q(M)$ of magnitude $M$
such that $0 \le q(M) \le 1$. The intensity $\lambda (M)$ for actually
observed magnitude frequency is described by $\lambda (M) = \lambda_{0}(M)
q(M)$, corresponding to thinning or random deletion. An example of the
detection rate function is the cumulative of Gaussian distribution or the
so-called error function $q(M) = \operatorname{erf}\{ M| \mu,\sigma \}$. The parameter
$\mu$ represents the magnitude at which earthquakes are detected at a rate
of 50\%, and $\sigma$ represents a range of magnitudes in which earthquakes
are partially detected. Let a data set of magnitudes $\{ (t_{i},M_{i}); i =
1, \ldots,N\}$ be given at a period immediately after the main shock. Assume
that the parameters are time-dependent during the period such that
%
\begin{equation}\label{eq4}
\lambda (t,M) = \frac{10^{a + b(M_{0} - M)}}{(t + c)^{p}}q \bigl\{ M|\mu (t),\sigma \bigr\}
\end{equation}
with an improving detection rate $\mu (t)$. An additional parametric
approach proposed by \citet{OMIetal} uses the state--space representation
method for real-time forecasting within the 24~h period.

\subsubsection{Epidemic-type aftershock sequence (ETAS)
model}\label{sec2.2.3}

\ The epidemic-type aftershock sequence\break  (ETAS) model describes earthquake
activity as a point process (\cite{OGA86a}, \citeyear{OGA88}) and includes the O--U law for
aftershocks as a descendant process. This model assumes that the background
seismicity is a stationary Poisson process with a constant occurrence rate
or number of earthquakes per day, $\mu$. The conditional intensity function
of the process is described by
%
\begin{equation}\label{eq5}\quad
\lambda_{\theta}(t|H_t)=\mu+\sum_{\{i:t_{i}<t\}}\frac{K}{(t-t_{i}+c)^{p}}e^{\alpha(M_i-M_{0})},
\end{equation}
%
where $H_{t}= \{(t_{i}, M_{i}); t_{i} < t\}$ is the history of the
occurrence times and magnitudes of earthquakes before time~$t$, and $M_{0}$
is a reference magnitude throughout the data; it can be a threshold
magnitude in case of general seismic activity or a main shock magnitude in
case of a single aftershock sequence. The parameters $K$, $\alpha$, $c$ and
$p$ are constants, and their detailed features are summarized and discussed
in Utsu, Ogata and Matsu'ura (\citeyear{UTSOGAMAT95}), for example. Here, in simulations and forecasting,
magnitude sequence is usually assumed to be independent and identically
distributed according to the G--R law (Section~\ref{sec2.2.1}) unless otherwise
modeled like in \citet{OGA89}.

We estimate the ETAS parameters by using the maximum-likelihood estimation
where the log-likeli\-hood function, or rigorously partial log-likelihood
(\cite*{C75}),
%
\begin{eqnarray}\label{eq6}
\log L (\theta; S,T)&=&\sum_{\{i; S<t_i<T\}}\log
\lambda_{\theta}(t_{i}|H_{t_i})
\nonumber
\\[-8pt]
\\[-8pt]
\nonumber
&&{}-\int^{T}_{S}\lambda_{\theta}(t|H_{t})\,dt
\end{eqnarray}
%
is maximized with respect to the parameters $\theta = (\mu, K, c, \alpha,
p)$. Here, $\{(t_i, M_{i}), M_{i} \geq M_{c}; i = 1, 2,
\ldots\}$ are data from the period [$0, T$] consisting of occurrence times
and magnitudes of earthquakes above a thereshold $M_c$. Here, note that the magnitudes are exogenous
variables. The ETAS model is applied to data from the target time interval
[$S, T$]. The occurrence history $H_{S}$ during the precursor period [$0,
S$] is used for sustaining stationarity of the process after the time
$S$.

Then, the model's effectiveness in fitting an earthquake sequence can be
evaluated by comparing the cumulative number $N (S, t)$ of earthquakes with
the rate predicted by the model
%
\begin{equation}\label{eq7}
\Lambda(S,t)=\int_{S}^{t}\lambda(u|H_{u})\,du
\end{equation}
%
in the time interval
$S<t<T$. If
earthquakes in the catalog are described effectively by the ETAS model, the
transformed time $\tau_{i}$ defined as $\tau_{i}=\Lambda(t_i)$, which include correction for
the O--U law decay, will be distributed according to the stationary Poisson
process, and the plot of the actual cumulative number of events versus
transformed time should be close to linear (\cite{OGA88}). The transformed
time $\tau_{i}$ is useful for
judging goodness of fit of the ETAS model because it assigns a visual check
of the fit to a stationary Poisson process. Anomalous seismicity, not
explained by the stationary ETAS model, will appear as systematic
deviations from this trend. An example of such an analysis will be
presented in Section~\ref{sec4.3} and Figure~\ref{f04}.

To predict in real time, the probability of\break  occurrence of future
earthquakes using the data\break  of earthquakes in the past, the ETAS model has
been used. For example, the ETAS model and\break  its space--time extensions (see
Section~\ref{sec2.2.4}) are reviewed in the next version on operational earthquake
forecast in California (Working Group\break  on California Earthquake
Probabilities, WGCEP,\break  \citeyear{WOR}).

\subsubsection{Space--time ETAS model}\label{sec2.2.4}

\begin{figure}

\includegraphics{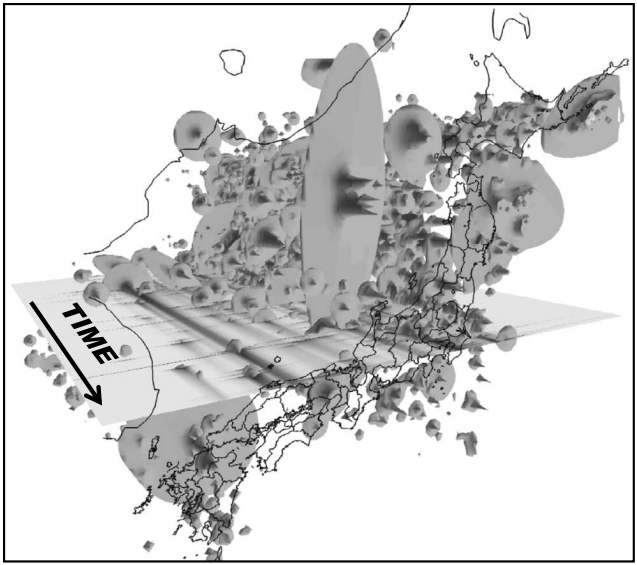}

\caption{Iso-surface plot of the estimated conditional intensity
function (\protect\ref{eq8}) of the space--time ETAS model (Ogata, \citeyear{OGA98}) to the JMA hypocenter
data of shallow earthquakes (depth $\leq 100$~km) of magnitude 5.0 or larger
from the period 1926--1995; in addition, Utsu's earthquake catalog for the 40
years period before 1926 (1885--1925) was used as the preceding occurrence
history of the space--time ETAS model. Space means longitude and latitude,
whereas the depth data are neglected.}\label{f02}
\end{figure}

The space--time ETAS model considers space--time occurrence rate at the
time $t$ and location ($x, y$), conditional on the occurrence history up to
time $t$, 
such that
%
\begin{eqnarray}\label{eq8}
&&\lambda (t,x,y|H_{t})\nonumber\\
&& \quad =\mu (x,y) \nonumber\\
&&\qquad{}+ \sum
_{j}^{t_{j} < t} \frac{K}{(t -
t_{j} + c)^{p}}
\nonumber
\\[-8pt]
\\[-8pt]
\nonumber
&&\hspace*{29pt}\qquad{}\times\Biggl[\frac{(x - \bar{x}_{j}, y - \bar{y}_{j})
S_{j}{x -\bar{x}_{j}\choose y - \bar{y}_{j}}}{e^{\alpha (M_{j} - M_{c})}}
+ d \Biggr]^{_{ - q}},\nonumber
\end{eqnarray}
where $S_{j}$ is a normalized positive definite symmetric matrix for
anisotropic clusters such that
%
\begin{eqnarray}\label{eq9}
&&(x,y)S(x,y)^{t}
\nonumber
\\[-8pt]
\\[-8pt]
\nonumber
&&\quad=\frac{1}{\sqrt{1-\rho^{2}}}\biggl\{\biggl(\frac{\sigma_2}{\sigma_1}\biggr)x^{2}
+2\rho x y +\biggl(\frac{\sigma_1}{\sigma_2}\biggr)y^{2}\biggr\}.
\end{eqnarray} 
Here, $(\bar{x}_j,\bar{y}_j)$ 
is an average location of earthquakes
that are placed in the same cluster as ($x_{j},y_{j}$). Both
$(\bar{x}_j,\bar{y}_j)$
and coefficients of $S_{j}$ for a selected
set of large earthquakes $j$ are identified by fitting a bivariate normal
distribution to spatial coordinates of the cluster occurring within a
square of $3.33 \times 10^{0.5M_{j} - 2}$~km side-length and
within $10^{0.5M_j-1}$ 
days after the large event of magnitude
$M_{j}$, according to \citet{UTSN1}; but I use 1~h in prediction stage. The
locations $(\bar{x}_j,\bar{y}_j)$
of all other events, including
cluster members, remain the same as the epicenter coordinates of the
original catalog; and they are associated with the identity matrix for
$S_{j}$, namely, $\sigma_{1} = \sigma_{2} = 1$ and $\rho = 0$. See Figure~\ref{f02}
for an illustrative view of the conditional intensity (\ref{eq8}). Further details
of the algorithm can be found in studies of Ogata (\citeyear{OGA98,OGA11aN8,OGA11N1}).

Although several alternative versions to the spatial factor given by the
bracket of (\ref{eq8}), as described by \citet{OGA98}, are available, the form in
(\ref{eq8}) fits best in terms of the Akaike information criterion (AIC; \cite*{A74}) for
Japanese earthquake data sets. All extensions of the temporal ETAS model
are referred to as space--time ETAS models (e.g., \cite*{NANetal12}).

\subsubsection{Hierarchical space--time ETAS model}\label{sec2.2.5}

When a region becomes wide or the number of earthquakes becomes
sufficiently large, spatial heterogeneity of seismicity becomes
conspicuous. For example, many studies have been conducted on regional
variation of seismicity-related parameters such as the $b$-value of the G--R
law and $p$-values of the O--U law (Utsu, \citeyear{UTS61}, \citeyear{UTSN1}; \cite*{M67}).

Regarding space--time ETAS models, the aftershock productivity $K$ may differ
significantly among locations, even if magnitudes of triggering earthquakes
are similar (see Section~\ref{sec2.2.2}). Moreover, the main shock--aftershock and
swarm-type clusters exhibit significantly different activity patterns.
Therefore, we applied an extension to the above space--time model to
earthquakes in the entire region for developing a hierarchical space--time
ETAS (HIST--ETAS) model, which is a space--time ETAS model in which
parameter values $\mu, K$, $\alpha$, $p$ and $q$ can vary depending on
location, such as $\mu (x, y),K(\bar{x}_{j},\bar{y}_{j}),\alpha
(\bar{x}_{j},\bar{y}_{j}),p(\bar{x}_{j},\bar{y}_{j})$
and $q(\bar{x}_{j},\bar{y}_{j})$.

\begin{figure}

\includegraphics{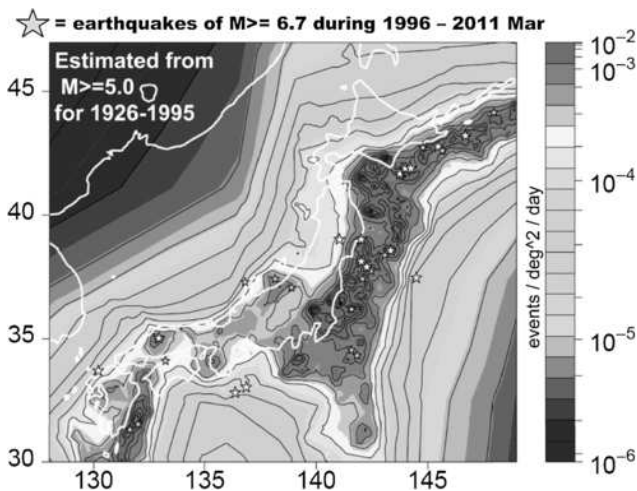}

\caption{The optimal solution of the $\mu$-values for background
seismicity of the space time ETAS model (\protect\ref{eq8}) in terms of minimum ABIC
priors. The model is estimated from the JMA data with earthquakes of M5.0
or larger for the target period 1926--1995. In addition, Utsu's earthquake
catalog for the 40 years period before 1926 (1885--1925) was used as the
precursory occurrence history of the space--time ETAS model. Contours are
equidistant in the logarithmic scale. Stars indicate locations of
earthquakes of magnitude 6.7 or larger that occurred during 1996--2009,
which mostly occurred in high background rates. Note however that there are several
large earthquakes occurred at very low seismicity rates, the issue of which lead to Section~\protect\ref{sec2.3}.}\label{f03}
\end{figure}

Thus, coefficients of parameter functions of the space--time ETAS model in
equation (\ref{eq8}) must be evaluated. Coefficients of each parameter function are
defined by values at epicenter locations of earthquakes and a number of points
on the region boundary. Hence, each function is uniquely defined by linear
interpolation of values at three nearest points (earthquakes) determined by
Delaunay tessellation that is constructed by all the earthquake locations
and additional points on the boundary of the region (see Figure~\ref{f01}).

For a stable optimal estimation, the freedom of coefficients of parameter
functions needs to be constrained to assign penalties against roughness of
the functions. The coefficients that maximize the penalized log-likelihood
are then sought, which is the equivalent of attaining the maximum posterior
distribution. Here, we adjusted the optimal prior function for the
parameter constraints in terms of the penalty function by an empirical
Bayesian method (\cite*{Aka80N1}). Further details can be found in the studies
of \citet{OgaKatTan03} and Ogata (\citeyear{OGA04N1,OGA11N1}). Figure~\ref{f03} shows the optimal
solution of background seismic activities $\mu (x, y)$, which appear useful
for long-term prediction of large earthquakes in and near Japan. Moreover,
stochastic declustering using the space--time ETAS model can make
realizations of background seismicity (Zhuang, Ogata and
  Vere-Jones, \citeyear{ZhuOgaVer02}; \cite*{ZHUetal05N2}; \cite*{BANOGA13}).

\subsection{Long-Term Probability Forecasts of Characteristic
Earthquakes}\label{sec2.3}

A characteristic earthquake is a repeating large earthquake that is
traditionally defined from paleoseismology observations. The estimation is
made by using recurrence times of a large earthquake on an active fault or
a particular seismogenic region on a plate boundary. The Earthquake
Research Committee of Japan (ERC, \citeyear{EAR01}) adopted the Brownian Passage Time
(BPT; Matthews, Ellsworth and Reasenberg, \citeyear{MATELLREA02}) renewal process, in which the inter-event
probability density function is given by
%
\begin{equation}\label{eq10}
f(x|\mu,\alpha)
=\sqrt{\frac{\mu}{2\pi\alpha^{2}x^{3}}}\operatorname{exp}\biggl\{-\frac{(x-\mu)^{2}}{2\mu\alpha^{2}x}\biggr\}.
\end{equation}
%
This equation considers the potential of further model extensions by useful
physical concepts in the elastic rebound theory, such as stress interaction
from neighboring earthquake ruptures. This physical concept will be
subsequently described in Sections~\ref{sec4} and \ref{sec5}.
BPT renewal process is based on
the following Brownian perturbation process:
%
\begin{equation}\label{eq11}
S(t)=\lambda t+\sigma W(t), \qquad t\geq0,
\end{equation}
%
which includes linearly increasing drift for stress accumulation and
diffusion rate $\sigma$. An earthquake occurs when the path $S(t)$ attains
the critical stress level $s_{f}$, and the accumulated stress is released
down to the ground state $s_{0}$ based on elastic rebound theory of
earthquakes (\cite{REI10}). Random fluctuations represent the transient
stress changes due to the effect of other earthquakes in close proximity
(see Section~\ref{sec4}). This model includes four parameters: the stress
accumulation rate $\lambda$, perturbation rate $\sigma$, failure state
$s_{f}$, and ground state $s_{0}$. If we assume that failure and ground
states $s_{f}$ and $s_{0}$, respectively, are constant, the interval of
earthquakes is independent and identically distributed with the BPT
distribution, in which parameters are related by $\mu = (s_{f} -
s_{0})/\lambda$ and $\alpha = \sigma /\sqrt{\lambda (s_{f} - s_{0})}$.

Because of very small sample size available from each fault, the mean
parameter $\mu$ has been estimated using methods other than the MLE. The
ERC (\citeyear{EAR01}) uses a common $\alpha$ value of 0.24 throughout Japan. This is
because better fit of the same $\alpha$ value was shown by the AIC
comparison than the different $\alpha$ estimates for respective active
faults, for a set of occurrence data with moderate sample sizes from four
active faults (ERC, \citeyear{EAR01}). Also, the ERC has estimated $\mu$ in two ways:
as the mean of past recurrence intervals and as expected intervals
estimated from the slip data of the fault plane. The latter estimate is
expressed by $T = U/V$, where $U$ is the slip size per earthquake and $V$
is the deformation rate per year, observed from the escarpment of the
fault. The ERC selects and applies one of these estimates for $\mu$ of each
active fault according to reliability of the data.

Alternatively, \citet{NOMetal11} propose following Bayesian estimation
procedure assuming a common prior distribution for fault segments
throughout Japan. Consider historical occurrence data $\mathbf{X}_{j} = \{
X_{i}^{j}; i = 1,2, \ldots,n\}$ in the $j$th segment of $m$ fault segments.
Consider a posterior density
%
\begin{eqnarray}\label{eq12}
&&\operatorname{posterior}(\mu_{j},\alpha_{j}|T_{j},
\phi_{\mu},\phi_{\alpha} )
\nonumber
\\[-8pt]
\\[-8pt]
\nonumber
&&\quad = L(\mu_{j},
\alpha_{j}|\mathbf{X}_{j}) \pi_{1}(
\mu_{j}|T_{j},\phi_{\mu} ) \pi_{2}(
\alpha_{j}|\phi_{\alpha} ),
\end{eqnarray}
where likelihood $L$ is based on the renewal process taking account of the
forward and backward recurrence times (\cite*{DalVer03}) and
$T_{j}$ is the above-mentioned geologically estimated slip deformation
ratio from slip data. Furthermore, the values of the
hyperparameters $\phi_{\mu}$ and $\phi_{\alpha}$ characterizing the prior
densities of $\mu \mbox{ and } \alpha$ are common to all considered fault
segments. We obtain their estimates by maximizing the integrated posterior
distribution
%
\begin{eqnarray}\label{eq13}
 &&\Lambda (\phi ) = \prod_{j = 1}^{m}
\int_{0}^{\infty} \int_{0}^{\infty}
\operatorname{posterior}(\mu_{j},\alpha_{j}|
\nonumber
\\[-8pt]
\\[-8pt]
\nonumber
&&\hspace*{140pt}{}T_{j},
\phi_{\mu},\phi_{\alpha} )\,d\mu_{j}\,d\alpha_{j},
\end{eqnarray}
where the subscript $j$ represents the $j$th segment of $m$ fault segments.
This maximizing procedure is called the Type II maximum likelihood method
(\cite*{G65}). The selection of the best combination of the prior
distribution factors and the optimal values of the hyper-parameters in (\ref{eq12})
are carried out to attain the smallest value of the Akaike Bayesian
information criterion (\textit{ABIC}; \cite*{Aka80N1}) that is defined by
$\mathit{ABIC} =\break  -2 \max_{\phi}\log \Lambda(\phi) +2\dim(\phi)$, where
$\dim\{\phi\}$ denotes the number of hyperparameters.

The most common forecast technique is a plug-in method, which is a
probability forecast that uses a distribution or conditional intensity
function with a parameter set to its estimated value, such as MLE. This
method works well if the estimation error is sufficiently small. However,
its predictive performance can be inadequate when the sample size is small.
Hence, ERC adopts the plug-in method only for $\mu$ 
and uses a
common $\alpha$ value of 0.24 throughout Japan. Alternatively, \citet{RvDD94}, Ogata (\citeyear{OGA99N2}, \citeyear{OGA02}) and \citet{NOMetal11} propose the Bayesian
prediction (\cite{Aka85})
%
\begin{eqnarray}\label{eq14}
 \tilde{h}(y|\mathbf{X})&= &\int_{0}^{\infty} \int
_{0}^{\infty} h (y|\mu,\alpha ) \bigl\{ 1 - F(y|\mu,
\alpha ) \bigr\}
\nonumber
\\[-8pt]
\\[-8pt]
\nonumber
&&\hspace*{39pt}{}\times\prod_{i = 1}^{n} f
(X_{i}|\mu,\alpha ) \,d\mu \,d\alpha,
\end{eqnarray}
where $F(y|\mu,\alpha )$ is the cumulative distribution of the
density $f(y|\mu,\alpha )$ in (\ref{eq10}), and $h(y|\mu,\alpha )$ is the hazard rate
function. This prediction is shown to provide a systematically better
performance in the sense of expected entropy criterion (\cite{Aka85}) than
the plug-in predictor in case of very small sample sizes of data 
(\cite{NOMetal11}).\vadjust{\goodbreak}

Bayesian framework can also be used when the occurrence times are
uncertain, especially when we are dealing with geological data (\cite*{OGA99N2}; \cite*{NOMetal11})
in addition to the magnitude dependent model (\cite{OGA02}) based on the time-predictable model (\cite{SHINAK80}).

\section{Practical Earthquake Forecasting}\label{sec3}

Probability gain refers to the ratio of predicted conditional probability
relative to baseline earthquake probability. As far as I know, most
probability gains of predictions are not very high, even relative to the
stationary Poisson process model. Therefore, predictions based on a single
anomaly data set alone are not satisfactory for disaster prevention because
baseline probability of a large earthquake itself is very small according
to the G--R law. Also, the BPT renewal process model has been applied to
active fault segments to estimate time-dependent probability on the basis
of the last earthquake and stress accumulation rate. In Californian,
probability gain showed an improvement of approximately 1.7 times over
Poisson process model predictions (\cite{JORetal11}).

The key for research progress in practical probability earthquake
forecasting is to use a multiple prediction formula (\cite{UTSN3}) such that
total probability gain is approximately equal to the product of individual
probability gains (\cite{AKI81}). The rate of probability gain that an
individual anomaly was actually a precursor to an earthquake may be
calculated as its success rate of the anomaly divided by precursor time
(\cite{UTSN3}). Success rate can only be determined from accumulation of
actual earthquake occurrences; and precursor time can be studied
experimentally and theoretically (\cite{AKI81}). In this section I review
important suggestions by \citet{UTSN3} and \citet{AKI81} and provide some
examples of causality modeling toward improved accuracy for probability
gain.

\subsection{Abnormal and Precursor Phenomena}\label{sec3.1}

The continuing pursuit of possible algorithms used to predict large
earthquakes should consider specific developmental patterns listed in the
seismic catalog. So far, an alarm-type method of earthquake prediction
(Keilis-Borok et al., \citeyear{KEIetal88}; \cite*{KEIMAL64}; \cite*{RUNetal02}; \cite*{SHEetal06}; \cite*{SOB01}; Tiampo, et al., \citeyear{TIAetal02})
based on seismicity patterns has
been operationally implemented, in which predictions\vadjust{\goodbreak} are conveyed to
seismologists through e-mail. Some predictions in each year are published
in an official document such as the Center for Analysis and Prediction,
State Seismological Bureau, China (\citeyear{Center90}). In addition, many papers
have been published on earthquake predictions. Some of these predictions
may be statistically significant against the stationary Poisson process
that is assumed for normal seismicity. Such alarm-type predictions have
been further evaluated by \citet{ZECZHU10}, \citet{JORetal11},
\citet{ZHUOGAN2} and \citet{ZHUJIA}.

Comprehensive studies of anomalous phenomena and observations of earthquake
mechanisms are essential for predicting large earthquakes with high
probability gains. However, it is difficult to determine whether detected
abnormalities are precursors to large earthquakes. Nevertheless, we aspire
to become able to say that the probability of occurrence of a large
earthquake, in a certain period and a certain region, has increased a
certain extent as compared with the reference probability. Therefore, it is
necessary to estimate uncertainty of the nature and urgency of abnormal
phenomena relative to their roles as precursors to major earthquakes. For
this purpose, it is necessary to study a large number of anomalous cases
for potential precursory links to large earthquakes. Thus, incorporation of
this information in the design of a prediction model for probability that
exceeds the underlying probability is important.

\subsection{Conditional Probability of an Earthquake for Multiple
Independent Precursors}\label{sec3.2}

As previously described, although an individual precursory anomaly is
insufficient for providing a forecast of an earthquake with a high
probability, forecasting probability can be enhanced if several anomalies
are simultaneously observed (Utsu, \citeyear{UTSN2,UTSN3,UTSN4}; \cite*{AKI81}). The
probability of an anomaly being a precursor of a large earthquake should be
estimated through comprehensive observations. Then, it provides medium- or
short-term probability forecasts depending on the time scale of enhanced
earthquake probability following the anomaly. For example, identification
of foreshocks (Section~\ref{sec3.4.2}) and seismicity quiescence (Section~\ref{sec5.1})
belongs to short- and medium-term forecasting, respectively.

Let us find the probability ${P}(E_{M} | A, B, C,\ldots, S)$ of
occurrence of an earthquake, with a magnitude greater than $M$ in a
specified area, under the condition that $N$ anomalies $A, B, C,\ldots,S$
appeared simultaneously. Assuming that anomalies are conditionally
independent on $E_{M}$ and the complement of $E_{M}$, \citet{AKI81} derived
the following equation of Utsu (\citeyear{UTSN2}, \citeyear{UTSN3}) using Bayes' theorem:
%
\begin{eqnarray}\label{eq15}
\hspace*{10pt}&&P(E_{M}|A,B,C, \ldots,S)\nonumber\\
&&\quad = \biggl[1+\biggl(\frac{1}{P_{A}}- 1\biggr) \biggl(\frac{1}{P_{B}} - 1  \biggr)
\biggl(\frac{1}{P_{C}} - 1  \biggr)\cdots
\\
&&\hspace*{77pt} {}\cdot\biggl(\frac{1}{P_{S}} - 1 \biggr) \Big/
\biggl(\frac{1}{P_{0}} - 1  \biggr)^{N - 1}\biggr]^{-1},\nonumber
\end{eqnarray}
where $P_{0} = P(E_{M})$, $P_{A} =  {P}(E_{M} | A)$, $P_{B} =
{P}(E_{M} | B),\break  \ldots, P_{S} = {P}(E_{M} | S)$. Note that
this formula can be written as a linear relation of logit functions of
probabilities [see equation (\ref{eq23}) in Section~\ref{sec3.4.2}]. These probabilities for a short time
interval become very small so that (\ref{eq15}) can be approximated by
%
\begin{equation}\label{eq16}
P(E_{M}|A,B,C, \ldots,S) \approx P_{0}
\frac{P_{A}}{P_{0}}\frac{P_{B}}{P_{0}}\frac{P_{C}}{P_{0}} \cdots \frac{P_{S}}{P_{0}}.\hspace*{-25pt}
\end{equation}

The above relation shows that for multiple independent precursors, the
conditional rate of earthquake occurrence can be obtained by multiplying
the unconditional rate $P_{0}$ with ratios of conditional probability to
unconditional probability $P_{0}$. Each ratio is defined as the probability
gain of a precursor. \citet{UTSN3} retrospectively reported a high
probability forecast of the 1978 Izu--Oshima--Kinkai earthquake of M7.0
using the multiple independent precursor formula. This is based on each
probability assessment of the anomalous phenomena consisting of uplift in
the Izu Peninsula, swarm, and a composite of a radon anomaly, anomalous
water table change and volumetric strain anomaly. Each of such
probabilities was not very high. \citet{AKI81} summarized the Utsu report and
further explained similar possible calculations for successful prediction
of the 1975 Haicheng earthquake of M7.3 in China by considering long-term,
intermediate-term, short-term and imminent precursory phenomena.

\subsection{Improving Probability Gains by Seeking Statistically
Significant Phenomenon}\label{sec3.3}

Here I would like to describe several point process models which can
enhance the probability gains. To examine whether certain abnormal
phenomena affect changes in the baseline rate of earthquake occurrences,
\citet{OgaAka82}, \citet{OGAAKAKAT82} and \citet{OgaKat86}
analyzed causal relationships between earthquake series from two different
seismogenic regions, $A$ and $B$. Let $N_{t}^{A}$ and $N_{t}^{B}$ be the number
of earthquakes above a certain magnitude thereshold  in the time interval ($0, t$) in
regions $A$ and $B$, respectively. Then, consider a model of the intensity
function of point process $N_{t}^{A}$ for earthquake occurrences in the
region $A$, conditional on the history of earthquake
occurrences $H_{t}^{A}$ and $H_{t}^{B}$ in both regions:
%
\begin{eqnarray}\label{eq17}
\hspace*{10pt}\lambda_{A}\bigl(t|H_{t}^{A},H_{t}^{B}
\bigr)&=&\mu + \sum_{j = 1}^{J}
a_{j}t^{j}+ \int_{0}^{t}
g(t - s)\,dN_{s}^{A}
\nonumber\\[-8pt]\\[-8pt]
&&{} + \int_{0}^{t} h(t - s)\,dN_{s}^{B},\nonumber
\end{eqnarray}
where the first two terms on the right-hand side of the equation represent
the Poisson process of a trend, the third term represents the cluster
component within region $A$ including aftershocks and swarms, and the last
term represents the effect of earthquake occurrences in region $B$.

Here, it must be noted that even if a significant correlation is observed
between the two series of events, it is insufficient from the standpoint of
prediction, and it is necessary to identify causality. Thus, we must
examine the opposite causality by interchanging $A$ and $B$ in equation
(\ref{eq17}). If both direction models hold, this process is mutually exciting
(\cite{Haw71}). Furthermore, the correlation between $A$ and $B$ regions
may be indirect such that activities in both regions may be affected by
additional factors, for which the trend term may be useful if the
polynomial can efficiently capture such an effect. According to our
analysis of seismicity in two seismogenic regions along the subducting
Pacific plate interface beneath the central Honshu, Japan, seismicity
causality was found as a one-way effect from the deeper to the shallower.
The maximum probability gain of the causal effect was several times the
average occurrence rate.

Similarly, we can examine the causal relationship from some observed
geophysical time series of $\xi_{s}$ (\cite*{OgaAka82}; \cite{OGAAKAKAT82}) as follows:
%
\begin{eqnarray}\label{eq18}
\hspace*{10pt}\lambda_{A}\bigl(t|H_{t}^{A},H_{t}^{\xi}
\bigr)&=& \mu + \sum_{j = 1}^{J}
a_{j}t^{j}+ \int_{0}^{t}
g(t - s)\,dN_{s}^{A}
\nonumber\\[-8pt]\\[-8pt]
&&{}+ \int_{0}^{t}
h(t - s)f(\xi_{s})\,ds.\nonumber
\end{eqnarray}

An example is the data of unusual intensities of ground electric potential,
which were observed in the vicinity of Beijing, China, during a 16-year
period beginning in 1982. The issue was whether or not these factors were
useful as precursors to strong earthquakes of $M\geq 4.0$. Electricity
anomalies could have been aftereffects of strong earthquakes. However, by
comparing the goodness of fit of models (\ref{eq17}) by AIC, anomalies were deemed
statistically significant as precursors to earthquakes (\cite*{ZHUetal05N1}). Moreover, the conditional intensity rate of declustered earthquakes
$M\geq 4.0$ or larger within a radius of 300~km from the Huailai
ground-electricity station was given by
%
\begin{eqnarray}\label{eq19}\qquad
 \lambda (t|H_{t}) &=& \mu + \int_{S}^{t}
h(t - s)\xi (s)^{a}\,ds
\nonumber
\\[-8pt]
\\[-8pt]
\nonumber
 &=& 0.00702 + \sum_{j = S}^{t}
0.000117 e^{ - 0.14 2(t - j)} \xi_{j}^{0.69}
\end{eqnarray}
(event/day) in the study of \citet{ZHUOGAN1}, in which successively
occurring $M\geq 4$
earthquakes within five days and
30~km distance were removed from the data to account for the self-exciting
effect in equation (\ref{eq18}). According to this model, the rate of
$M\geq 4$
earthquakes varies from a half to 10 times
the average occurrence of 0.0126 event/day.

Furthermore,
the time series of electric anomaly records were available from three other
stations near Beijing. If we assume that the four sets of the time series
are approximately independent, we may consider the following conditional
intensity rate by extending the multiple precursor in equation (\ref{eq16}):
%
\begin{eqnarray}\label{eq20}
\lambda_{A}\Biggl(t\Big| \bigcap_{m = 1}^{4} H_{t}^{m} \Biggr) \approx
\hat{\lambda}_{A}\prod_{m = 1}^{4}
\frac{\lambda_{A_{m}}(t|H_{t}^{m})}{\hat{\lambda}_{A_{m}}}
\end{eqnarray}
for the common region $A = \bigcap_{m = 1}^{4} A_{m}$ among four circular
regions $A_{m}$ of 300~km radii from the four stations. Retrospective total
probability gain varies in the range 1/10--100 times of the average
occurrence rate $\lambda_{A}$ in the common region (\cite*{ZHUOGAN1}).

Here, we considered declustered earthquakes near Beijing, but if we can
consider the original data and model that take the triggered clustering
effect into account (cf. Zhuang et al., \citeyear{ZHUetal05N1}, \citeyear{Zetal13}), the corresponding
model would become
%
\begin{eqnarray}\label{eq21}\qquad
\lambda \Biggl(t\Big| \bigcap_{m = 1}^{4}
H_{t}^{m} \Biggr) \approx \lambda_{0}(t|H_{t})
\prod_{m = 1}^{4} \frac{\lambda
(t|H_{t}^{m})}{\lambda_{0}(t|H_{t})}.
\end{eqnarray}

A similar but more general model is applied in foreshock forecasting
discussed in the following section (Section~\ref{sec3.4.2}).

Moreover, we examined periodicity, or seasonality, of earthquake
occurrences (Ogata and Katsura, \citeyear{OgaKat86}; \cite*{MaVere97}). Although
such issues have been frequently discussed in statistical seismology, it
was difficult to analyze correlations in the conventional method because
the clustering feature of earthquakes frequently leads to incorrect results
(Aki, \citeyear{A56}). On the other hand, we found it effective to apply statistical
models of stochastic point processes that incorporate a clustering
component; further details can be found in the study of \citet{OGA99N1}, and
the references therein. These models can also be applied to examine whether
or not various geophysical anomalies are statistically significant as
precursors of an approaching large earthquake:
%
\begin{eqnarray}\label{eq22}
\lambda_{\theta}(t|H_t)&=&\mu+\sum_{j=1}^{J}a_j t^j\nonumber
\\
&&{}+\sum^{K}_{k=1}\biggl\{c_{2k-1}\cos\frac{2\pi k t}{T_0}+c_{2k}\sin\frac{2\pi
kt}{T_0}\biggr\}
\\
&&\qquad{}+\int^t_0 g(t-s)\,dN_s.\nonumber
\end{eqnarray}

From estimated amplitudes of the one-year periodic term with $T_{0} =
365.24$ days, it is evident that probability gains vary around the average
occurrence rate of corresponding $M\geq 4.0$ and $M\geq 5.0$ earthquakes. More extensive
studies were reported by \citet{MAT86} by using the above model, in
which the trend-term (first two terms) was used for artificial
nonstationarity due to an increasing number of observed earthquakes in the
long-term global catalog. He detected \mbox{periodic} effects for many
mid-latitude seismic inland regions, whereas the seasonality was rarely
observed in tropical seismic regions and ocean seismogenic zones.
Correlations with precipitation variations were common in these results and
are most probably due to pore fluid pressure changes in faults (see Section~\ref{sec4.1} for the physical mechanism). An extension of the above periodicity
model, reported by \citet{IWAKAT06}, includes a combination of
(lunar) synodic and semi-synodic periods to examine whether or not and how
certain seismicity is affected by Earth tides. Statistical models,
applications to validate data from the earthquake-induced phenomena and
their references were reviewed by \citet{OGA99N1}.

\subsection{Probabilistic Identification of Foreshocks}\label{sec3.4}

The study of foreshocks should lead to a short-term forecasting. Although a
considerable number of foreshocks are observed, most are recognized after
occurrence of a large earthquake. Nevertheless, when earthquakes begin to
occur in a local region, its residents should determine whether or not such
movement is a precursor of a significantly larger earthquake. The
probability of foreshock type can be determined statistically from the data
of ongoing earthquakes in a particular region. Moreover, by using composite
identification data of magnitude sequence and degree of hypocenter
concentration, the probability gain of prediction is heightened.

\subsubsection{Working definitions for foreshock
discrimination}\label{sec3.4.1}

When an earthquake of M4.0 or larger occurs, it must first be determined
whether or not the movement is a continuation of nearby earthquakes.
Precisely, the connection to past earthquakes is determined by the
single-link clustering (SLC) algorithm of Frohlich and Davis (\citeyear{FD90}).

The largest earthquake in a cluster is designated as the main shock.
Pre-shocks refer to all earthquakes preceding the main shock of a cluster.
All pre-shocks in a cluster become foreshocks when the magnitude gap or
magnitude difference between the largest pre-shock and main shock is 0.45
or greater. If the magnitude gap is smaller than 0.45, the cluster with
pre-shocks is defined as a swarm. An additional type of cluster is the main
shock--aftershock type, in which the main shock occurs first in the
cluster. A magnitude gap of 0.45 or larger between the main shock and
largest pre-shock occurs in less than approximately 20\% of pre-shock
clusters in Japan. This 0.45 borderline of foreshock- and the swarm-types
has been determined by a trade-off between achievement of a larger
magnitude gap, which results in better discrimination of the foreshock, and
a greater number of foreshock clusters, which results in better statistics.
Here, to characterize these features, we note that clusters of
foreshock-type exclude main shock and subsequent aftershocks, whereas other
cluster types include all events in each cluster. This designation is made
because real-time recognition of the main shock, which is preceded by
foreshocks, is easy owing to the large magnitude gap, whereas main shocks
of other cluster types are difficult to recognize until the end of the
cluster.

\subsubsection{Probability forecast by discrimination of
foreshocks}\label{sec3.4.2}

Using the location ($x, y$) of the first earthquakes from clusters or
isolated single\vadjust{\goodbreak} earthquakes, by the empirical Bayesian logit model, Ogata, Utsu and Katsura (\citeyear{OGAUTSKAT96}) obtained a map $\mu (x, y)$ of a probability that the
earthquake will be a foreshock of a forthcoming main shock. Such
probability varies from 1\% to 10\% with an average of 3.8\% throughout
Japan. Probability forecasts using this map have been conducted from
January 1994 to April 2011, and their performances have been demonstrated
by \citet{OGAKAT12}.

Multiple earthquakes occurring in a cluster provide more effective forecast
updates, and certain statistics within the cluster are useful for
discriminating foreshocks. Ogata, Utsu and Katsura (\citeyear{OGAUTSKAT96}) revealed that distances
between foreshocks in time and space are statistically shorter than those
between earthquakes in clusters of other types. Moreover, increasing
magnitudes enhance the probability of foreshocks. In the following model,
we devise foreshock probability by using such statistics for prospective
forecasting of main shocks.

Suppose that multiple earthquakes occur in a cluster $c$. Then, we consider
time differences $t_{i,j} = t_{j} - t_{i}$ (days), epicenter separations
\[
r_{i,j} = \sqrt{(x_{j} - x_{i})^{2}\cos^{2}\theta_{i,j} + (y_{j} -
y_{i})^{2}}\mbox{ km},
\]
 where $\theta_{ij}$ represents the mean latitude of
earthquakes $i$ and $j$, and magnitude differences $g_{ij} = M_{j} - M_{i}$
between earthquakes $i$ and $j\ (i < j)$. On the basis of a comparative
study of these statistics (Ogata, Utsu and Katsura, \citeyear{OGAUTSKAT95,OGAUTSKAT96}), we standardized them
into a unit interval. Specifically, time difference was transformed by
$\tau = \log(100t)/\log(3000)$ for $0.1 \leq t \leq30$ days; otherwise, 0
and 1 were set for $t \leq 0.1$ and $t \leq 30$, respectively. Epicenter
separation was transformed by $\rho = 1 - \exp\{-\min\{(r, 50)/20\}$~km.
Finally, magnitude difference was transformed by $\gamma = (2/3)
\exp\{g/\sigma_{1}\}$ and $\gamma = 2/3 + (1/3)\{1 - \exp(g/\sigma_{2})\}$
for $g \leq 0$ and $g > 0$, respectively, where $\sigma_{1} = 0.6709$ and
$\sigma_{2} = 0.4456$ (km).

Suppose that, at the current time, $c|n$ shows the stage where the
$n$th earthquake ($n = 2, 3, 4, \ldots, \#c$) has occurred in a cluster
$c$, where $\#c$ is the number of all earthquakes in the cluster $c$. We
propose the forecasting probability $p_{ c| n}$ by using the following
logistic transformation: Set $f = \operatorname{logit} p = (1 - p)/p$, or $p =
1/(1 + e^{f})$; then,
%
\begin{eqnarray}\label{eq23}
\operatorname{logit}p_{c|n}
&=&\operatorname{logit}\mu(x_1,y_1)\nonumber
\\
&&{}+\frac{1}{\#(i<j\leq n)}
\nonumber\\\\[-16pt]
&&\hspace*{10pt}{}\times \sum_{i<j\leq n}\Biggl(\mu_0+
\sum_{k=1}^3 b_k\gamma^k_{i,j}\nonumber
\\
&&\hspace*{30pt}\qquad{}+\sum_{k=1}^3c_k\rho^{k}_{i,j}+\sum_{k=1}^3 d_k
\tau_{i,j}^k\Biggr).\nonumber
\end{eqnarray}
Here, the first term $\mu (x_{1}, y_{1})$ indicates the probability that
the first earthquake in the cluster is a foreshock, and the second term
indicates the sample mean of weighted polynomials of transformed variables
defined among all cluster members up to the time of forecasting. Here, the
factor $\#(i < j \leq n)$ is the number of pairs in the first $n$ members
of the cluster $c$.

It must be noted that interactions between the normalized statistics were
not selected in equation (\ref{eq23}) by the AIC comparison; namely, independency
for the formula (\ref{eq15}) is shown between the statistics of time intervals,
epicenter separations and magnitude differences. Such conditions can be
extended for a case in which the factors are dependent by considering
higher order of polynomials; however, the linear factor in equation~(\ref{eq23})
represented the best fit in this case according to AIC (Ogata, Utsu and Katsura,
\citeyear{OGAUTSKAT96}).

Probability forecasts that use prediction equation~(\ref{eq23}) have been conducted
from January 1994 to April 2011, and their performances have been evaluated
by \citet{OGA11aN8} and \citet{OGAKAT12}. Therefore, these forecasts are expected to be
applied for practical use in real time in the near future.

\section{Incorporating Physical Mechanisms of Earthquakes}\label{sec4}

\subsection{Earthquake Dynamics and Interactions}\label{sec4.1}

The earth crust and upper mantle lithosphere can be approximately considered as an
elastic body. These become distorted under stress, which increases steadily
in a particular direction. Fault planes are cracks within the lithosphere
or plate boundary interfaces. Earthquakes occur through distortion of
subsurface rocks and both sides of the fault plane moving out of alignment.
The earthquake location listed in hypocenter catalogs is location at which
a fault displacement started, and \mbox{earthquake} magnitude represents eventual
size of the displacement. Moreover, some catalogs record the orientations
and slips of fault planes of relatively large earthquakes.

For each fault plane, stress tensor in lithosphere is decomposed into two
perpendicular components. The shear stress acts in a direction parallel to
fault shifting, and the normal stress acts perpendicular to the fault
plane. The orientation of each fault plane determines shear and normal
stress, which define Coulomb failure stress (CFS):
%
\begin{eqnarray}\label{eq24}\quad\!\!
\quad\operatorname{CFS} &=& (\mbox{Shear Stress}) - (\mbox{friction
coefficient})
\nonumber
\\[-4pt]
\\[-12pt]
\nonumber
&&{} \times (\mbox{Normal Stress} - \mbox{pore fluid
pressure}).
\end{eqnarray}
CFS increases at a constant rate over time. When CFS exceeds a particular
threshold, the fault slips dramatically (an earthquake). Then, the stress
reduces to a certain value and accumulates again over decades to result in
large earthquakes on plate boundaries, and over thousands of years to
result in slips on inland active faults (see Section~\ref{sec2.3}).

When an earthquake occurs, the displacement of the source fault causes
sudden Coulomb stress changes ($\Delta$CFS) on the peripheral receiver
faults. The $\Delta$CFS of each receiver fault plane either decreases or
increases depending on its orientations relative to the slip angles of the
source fault. On the faults with increased $\Delta$CFS, earthquakes occur
earlier than expected, whereas on those with decreased $\Delta$CFS,
forthcoming earthquakes are delayed. When faults of similar orientations
dominate a region, either seismic activation or quiescence is expected in
the region.

In equation (\ref{eq24}), the pore fluid pressure of the gap fault related to CFS
is generally a constant. However, its changes may be important. For
example, pressure changes in the fluid magma gap affect swarm activity in
volcanic areas (Dieterich, Cayol and
Okubo, \citeyear{DIECAYOKU00}; Toda, Stein and Sagiya, \citeyear{TODSTESAG02}). In addition,
earthquakes can be induced through increased pore fluid pressure in a fault
system (\cite{HAIOGA05}; \cite{THM13}), which is occasionally due to heavy
rainfall or shaking of the earth crust owing to propagated strong seismic
waves; the latter causes dynamic triggering (Steacy, Gomberg and
Cocco, \citeyear{STEGOMCOCN1}, and
papers included in the same volume). Relevantly, the seasonal nature of
seismicity or annual periodicity has been discussed in Section~\ref{sec3.3} [cf.~equation (\ref{eq22})].
Moreover, the ETAS model applications to seismicity changes that were
induced by dynamic triggering or injection of water were reported (Lei et
al., \citeyear{LEIetal08}, Lei, Xie and Fu, \citeyear{LEIXIEFU11}).

\subsection{Predicting Seismicity in the Peripheral Area by Abrupt Stress
Changes}\label{sec4.2}

To explain earthquake induction or suppression of seismicity, it is useful
to determine whether or not $\Delta$CFS was due to a rapid faulting event
that caused an earthquake. When a large earthquake occurs, low-frequency
seismic waves and global positioning system (GPS) crustal displacement are
observed. From such observations, source fault mechanisms of the earthquake
can be solved, such as size, orientation and vector of the fault slip. Such
source parameters are input into a computer program designed by \citet{OKA92} to
calculate $\Delta$CFS in a receiver fault system on the basis of
source fault data. Thus, studies on induction of earthquakes, based on
$\Delta$CFS, have become popular. Special issue volumes on this subject
have been edited by \citet{HAR98} and \citet{STEGOMCOCN1}.

For example, \citet{OGA04N2} examined regional $\Delta$CFS in southwestern
Japan by analyzing M7.9 Tonankai and M8.1 Nankai earthquakes in 1944 and
1946, respectively. Conventionally, some seismic quiescence in this period
was either considered as a genuine precursor or suspected as an artifact
because of incomplete detection of earthquakes during the Second World War.
Positive and negative $\Delta$CFS correlated strongly with seismic
activation and quiescence, respectively. In particular, this study
classified seismicity anomalies into pre-seismic, coseismic and
post-seismic before, during and after massive earthquakes, respectively.
These scenarios may be helpful in interpreting seismicity in western Japan
prior to occurrences of expected subsequent large earthquakes along the
Nankai Trough.

\subsection{Physical Implication of the ETAS Model and
Seismicity}\label{sec4.3}

In general, interactions among earthquakes are fairly complex. Once an
earthquake occurs in a particular location, CFS of the fault system
adjacent to the source fault is considerably increased, and many
earthquakes are induced. Traditionally, these earthquakes are called as
aftershocks. Some of them are induced outside the aftershock region; these
are also called as off-fault aftershocks, or aftershocks in a broad sense.
Significant changes in stress result in many aftershocks; even small
changes can induce aftershocks to some extent. Furthermore, any aftershock
can change stress, too, causing their aftershocks. Because of such complex
interactions among invisible fault segments in the crust, detailed
calculations of such stress changes are difficult and impractical.

Therefore, statistical models designed to describe the actual macroscopic
outcome of these stress interactions are required. For example, the ETAS
model in equation (\ref{eq5}), which consists of empirical laws of aftershocks,
quantifies dynamic forecasting of induced effects. By fitting to the
selected data from the catalog earthquake, this ETAS model determines the
parameters by the maximum-likelihood method. Thus, prediction of
earthquakes conforming to regional diversity is possible.

On the other hand, the friction law of Dieterich (\citeyear{Die94}), which was developed on
the basis of rock fracture experiments with controlled stress, can be
linked to statistical laws of earthquake occurrences. In
particular, this law reproduces temporal and spatial distribution of the
attenuation rate of aftershocks, such as that determined by the O--U law in
equation (\ref{eq2}). However, because of seismicity diversity, predictions
adapting well to development of seismicity appear to be difficult.

Seismicity anomalies can hardly be detected by observing conventional plots
of earthquake series because they show a complex generation due to
successive occurrences of earthquakes or clustering. The clustering nature
is also the main difficulty for traditional statistical test analysis.
Complexity due to the clustering feature creates difficulties in revealing
anomalies of seismicity caused by slight stress changes, hence, various
anomalous signals are missed.

\begin{figure*}

\includegraphics{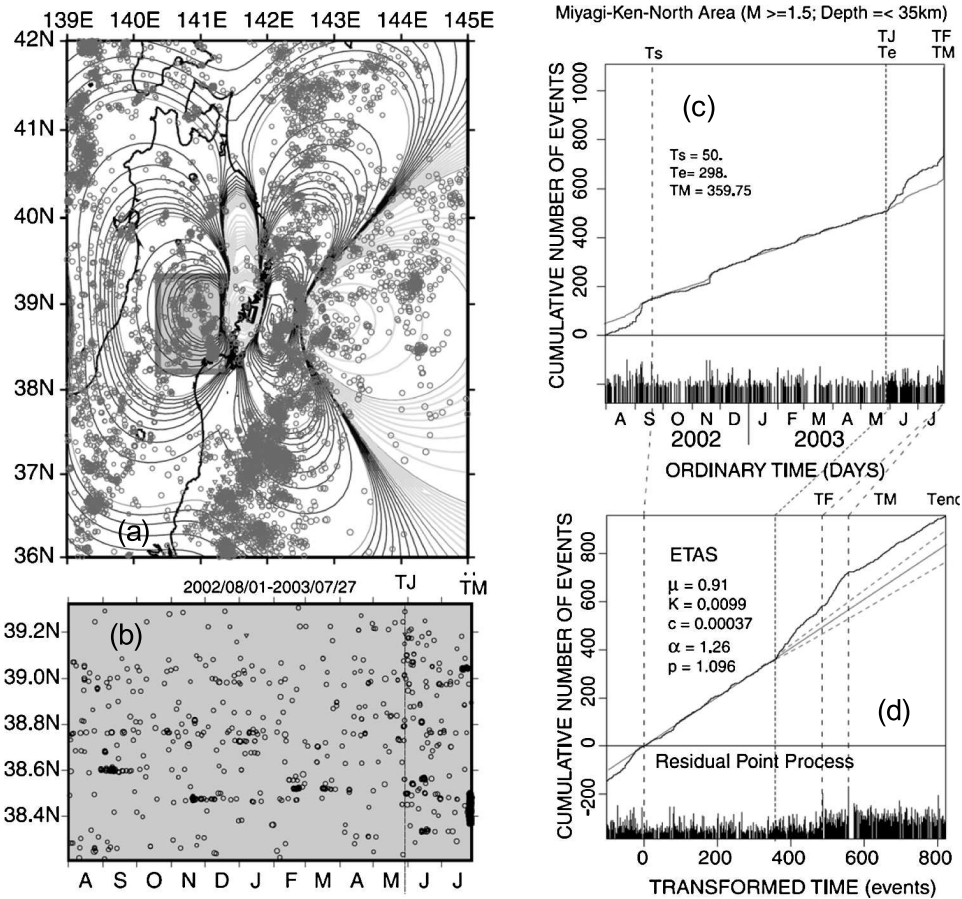}

\caption{Shallow earthquakes in the inland rectangular region in
\textup{(a)} were analyzed from August 2002--July 2003 to investigate the effect of
the M7.0 earthquake in 2003. CFS increments of this region took the largest
values transferred from the M7.0 earthquake that is shown by the small
rectangle fault located at (141.7{\textdegree}\textup{E}, 38.8{\textdegree}\textup{N}) at a
depth of 71~km. Epicenters and latitude versus time are shown in \textup{(a)} and \textup{(b)}, respectively. The
occurrence time of the M7.0 earthquake is shown in each panel as the
vertical dotted line indicated by TJ. The ETAS model was fitted to the data
from the target period (TS, TJ). The panel \textup{(c)} shows cumulative numbers and
magnitude versus ordinary time; and the panel \textup{(d)} shows these values
against the transformed time determined in equation (\protect\ref{eq7}) by the estimated
ETAS model. The black and grey cumulative function in panels \textup{(c)} and \textup{(d)}
show the empirical cumulative function (step function) and the theoretical one (curve and straight line)
estimated and then predicted by the ETAS model, respectively. At the
southeastern corner of the inland rectangular region, a large M6.2
earthquake (occurrence time indicated by TM) and its largest M5.5 foreshock
(indicated by TF) occurred on July 26, 2003. The panel \textup{(d)} shows that the
foreshock activity was more active than was expected, which is seen from the
steepest slope of the cumulative function in \textup{(d)}. In contrast, the
aftershock activity of the M6.2 earthquake, during the period TM-Tend,
appears to be similar to predicted rates in \textup{(d)}. Dotted parabola-like
envelope curves show twofold standard deviations (95\% error bands) of
cumulative numbers of the transformed time.}\vspace*{12pt}\label{f04}
\end{figure*}

Therefore, some seismologists have devised various declustering methods
that include only isolated and largest earthquakes in a clustering group or
the main shock, and other earthquakes are excluded. On the basis of
declustered data, statistical significance of seismic quiescence was tested
against the Poisson process. Occasionally, however, analysis results depend
on the choice of criteria of the adopted declustering algorithm (Van~Stiphout, Zhuang and Marsan, \citeyear{VAN12}). Hence, results could be due to artificial
treatment. In addition, declustering methods result in a significant loss
of information because they discard a large amount of data from the
original catalog.

The ETAS model uses original earthquake data without declustering. As
mentioned in Section~\ref{sec2.2.3}, the ETAS model is a point process model
configured to conform to empirical laws derived from various studies such
as aftershocks in Japan and the time evolution of seismicity rate. Because
regional characteristics of earthquake occurrences can be captured and
considered as typical seismicity in this model, it has been accepted by
seismologists as a standard model of ordinary seismicity. The ETAS model is
used as a ``barometer'' for detecting significant deviations from normal
activities as demonstrated in Figure~\ref{f04} in Ogata (\citeyear{OGA05a}).

\section{Seismicity Anomalies for Intermediate-Term
Forecasts}\label{sec5}

\subsection{Seismicity Quiescence Relative to the ETAS
Model}\label{sec5.1}

The deviation of actual cumulative number of earthquakes is measured
relative to the theoretical cumulative function of the earthquake that
serves as an indefinite integral in time for the predicted rate function
(\ref{eq7}) of the ETAS model. Relative quiescence occurs when actual earthquake
occurrence rates are systematically lowered in comparison with the
predicted incidence that is determined by the ETAS model (\cite{OGA92}).
Relative quiescence lasting for many years was observed in a broad region
before great earthquakes of M8 class and larger occurred in and near Japan
(\cite*{OGA92}, \citeyear{O06b}). Similar phenomena were observed before
M9-class large earthquakes in other regions of the world.\looseness=1

Since 2001, I have reported 25 agendas of various seismicity anomalies and
forecasting proposals in Japan at the Coordinating Committee for Earthquake
Prediction of Japan (CCEP). Except the agenda that reported seismic
quiescence of aftershock activity before the largest aftershock (see
Section~\ref{sec6.2} for detail), all were ex-post analysis report; the agendas were
summarized in \citet{OGA099}. In addition, among 76 aftershock cases in Japan
that I have investigated, relative quiescence was observed in 34 (see
\cite*{OGA01}, and its appendix for details of the case studies). Moreover,
Section~\ref{sec5.4} includes a discussion on the manner in which results of this
aftershock study will be used for space--time probability prediction of a
neighboring large earthquake with a size similar to that of the main
shock.\looseness=1

Here, I will note the results on the aftershock research of inland
earthquakes of M6.0 or larger in southwestern Japan that occurred 30 years
before and after the M8.1 Nankai earthquake in 1946. Among six earthquakes
which occurred before 1946, relative quiescence was observed in five
aftershock sequences. In contrast, among seven earthquakes after 1946,
relative quiescence was not observed in six aftershock sequences, and these
aftershock sequences were on track as expected. Since the ERC forecasts the
next large earthquake for the next 30 years 60--70\% (see Section~\ref{sec2.3},
also see Ogata, \citeyear{OGA01} and \citeyear{OGA02}), it would be worthy to monitor recent and
future aftershock activity of similar large inland earthquakes.

\begin{figure*}
\centering
\begin{tabular}{@{}cc@{}}

\includegraphics{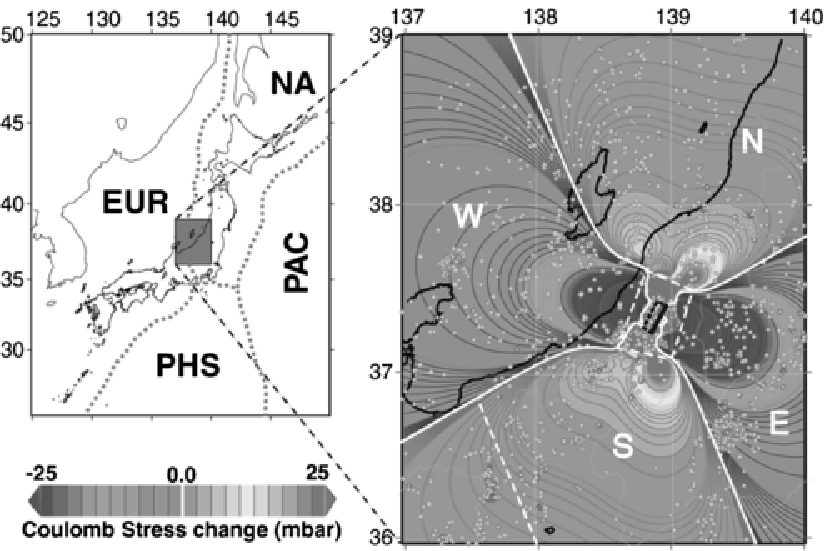}
 & \includegraphics{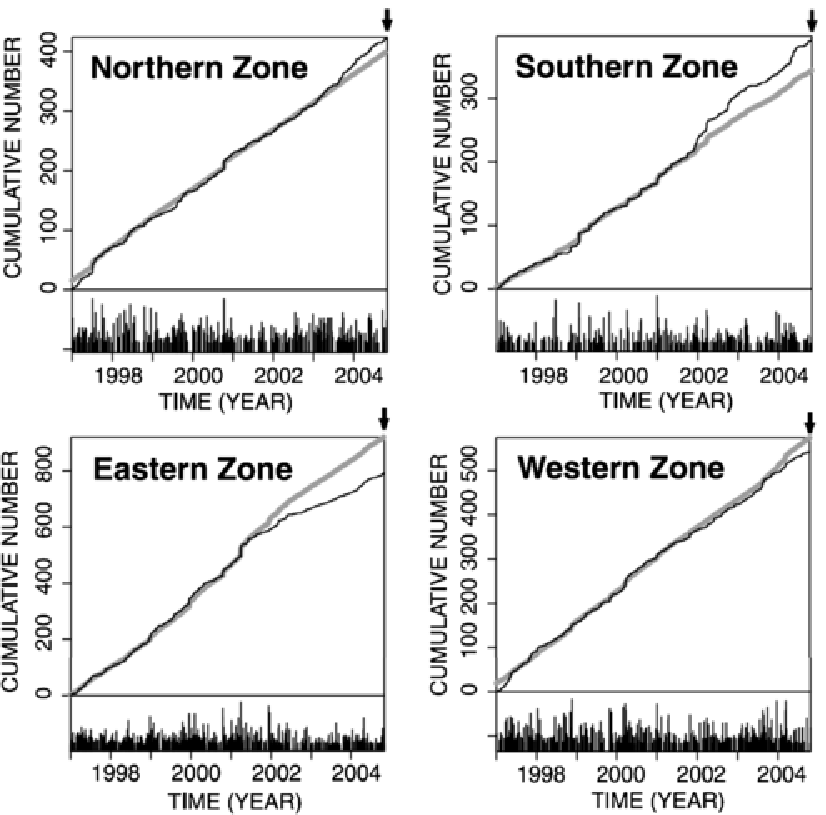}\\
\footnotesize{(a)} & \footnotesize{(b)}\\
\end{tabular}
\caption{\textup{(a)} Seismic activity for the eight years period before the M6.8
earthquake of 2004. Dominating fault slip orientations of earthquakes in
this area are similar to those of the main shock and its aftershocks. The
small black rectangle in the center of the magnified geographical map shows
the main shock fault model determined by the GPS observations. The regions
of thin and thick contours show positive and negative Coulomb failure
stress (CFS) increments, respectively, assuming that slow slips in the
source have occurred for some time. These define four subregions N, S, W
and E used for the following ETAS analysis.
\textup{(b)} The four panels show the empirical cumulative curve
(thin black) of the sequence of earthquakes of magnitude 2 or larger in
each of the four divided subregions from 1997 until the M6.8 earthquake
(downward arrows). Thick gray curves show estimated and predicted
cumulative functions before and after each change-point time, respectively.
Activation and quiescence relative to those predicted by the ETAS model
agrees with increase and decrease in CFS, respectively.}\label{f05}
\end{figure*}

\subsection{Aseismic Slip, Stress Change and Seismicity
Anomalies}\label{sec5.2}

Since a dense GPS observation network was established in Japan, aseismic
fault motions or slow slips that could not be detected by seismometers have
been successively identified in the plate boundary regions. We can take
occurrence of such motion into account in discussing the relationship
between seismicity anomalies (quiescence or activation) and stress
changes.

Specifically, it can be assumed that slow slips on a focal fault or its
adjacent part have occurred during a particular period. Then, depending on
dominating orientations of receiver faults neighboring the focal fault, CFS
could decrease or increase. Accordingly, we expect that seismicity there
decrease or increase relative to the expected occurrence rate by the ETAS
model. Such seismicity anomalies are revealed before some recent large
earthquakes (Ogata, \citeyear{OGA05b}, \citeyear{OGA07}, \citeyear{OGA10N1}, \citeyear{OGA11N2}; and Kumazawa, Ogata and Toda, \citeyear{KUMOGATOD10}).
See Figure~\ref{f05} for an example. By assuming slow slip on the source fault, the
peripheral regions were classified as either of the increasing or
decreasing CFS. Then, each region can theoretically correspond to an area
that either promoted or suppressed seismicity. Such anomaly patterns of
seismicity relative to the ETAS model (\ref{eq5}) are in good agreement with those
of CFS increment.

\subsection{Variation of Local Stress Deduced from Spatio-Temporal
Variation of Aftershocks}\label{sec5.3}

Local anomalies occur in space--time locations of most of the aftershocks.
To elucidate these anomalies, we firstly apply the O--U formula (\ref{eq2}) for
aftershock decay to data of occurrence times and convert these times by the
estimated theoretical cumulative function (\ref{eq7}). We then examine whether
space--time coordinates on a projected line, such as longitude and latitude,
against the converted time remained temporally uniform or not. If
nonuniformity in a certain portion of space--time conversion is observed,
this implies discrepancies between theoretical and actual aftershock
occurrences in such a place. Several possible scenarios for such
discrepancies are offered: Secondary aftershocks that follow a large
aftershock are obvious once seen as a cluster. Such a cluster shows traces
of a new local rupture to extend the peripheral portion of the fault of the
main shock. Moreover, when a nonuniform portion other than the secondary
aftershocks is observed, it is crucial for us to explore the reasons.

Based on recent accurate aftershock location data, \citet{OGA10N2} revealed
local relative quiescence and activations; these can occur associated with
post- or pre-slips of a large aftershock. These anomalies were
systematically investigated assuming that they were related to changes in
the CFS rate. In addition, assuming several scenarios of stress changes due
to slow slips, \citet{OGATOD10} and \citet{OGA10N2} performed
simulations to reproduce seismicity anomalies of relative activation and
quiescence within aftershocks on the basis of the rate/state friction law
of \citet{Die94}.

\subsection{Space--Time Probability Gain of a Large Earthquake Under
Relative Quiescence of Aftershocks}\label{sec5.4}

The probability of relative quiescence being precursor to a large
earthquake must be evaluated with their likely time and location. Because
these involve many conditions and a number of parameters, they cannot be
easily stated. However, by statistical studies of aftershock sequences in
Japan (Ogata, \citeyear{OGA01a}), what I can say about a probability gain that a large
earthquake will occur is as follows: First, if a large earthquake occurred
in a particular location, the probability per unit area that another
earthquake of similar magnitude will occur in the vicinity is greater than
that which will occur in a distant area. This is the result of simple
statistics regarding the self-similarity feature (inverse-power law
correlations), and also physically suggests that the neighboring earthquake
will be more probably induced by a sudden stress change on the periphery
because of the abrupt slip of the earthquake. Moreover, if aftershock
activity becomes relatively quiet, it becomes more likely that large
aftershocks will occur around the boundary of the aftershock area.
Furthermore, if relative quiescence lasts for a sufficiently long time more
than a few months, the probability that another earthquake of similar
magnitude will increase within six years in the vicinity of the aftershock
area within 200~km distance.

\section{Seismicity and Geodetic Anomalies}\label{sec6}

\subsection{Aseismic Slip and Crustal Deformation}\label{sec6.1}

The Geological Survey Institute (GSI) of Japan compiles daily geodetic
locations of global positioning system (GPS) stations throughout Japan, and
baseline distances between GPS stations can be calculated from data in the
GPS catalog. The geodetic time series show that contraction or extension of
the distance between stations is basically linear with time because the
subducting plate converges with constant speed. However, several years
prior to large inland earthquakes of M7 class, the time series of the
baseline distance variation around the fault was observed with systematic
deviation from a linear trend (Ogata, \citeyear{OGA07}, \citeyear{OGA10N1}, \citeyear{OGA11N2};
Kumazawa, Ogata and Toda, \citeyear{KUMOGATOD10}; GSI, \citeyear{GSI2009}). Each deviation of these baselines was consistently
explained by slow slips on the earthquake source fault or on its down-dip
extension. These results were due to post-hoc analysis based on knowledge
of the source fault obtained by coseismic displacement.

From a predictive perspective, it is highly desirable to estimate such a
fault slip in near real-time to that of occurrence. So far, several
estimates of sufficiently large slips on plate boundaries have been
obtained from GPS records by inversion analysis. GSI has regularly reported
such estimates of coseismic, post-seismic and large-size habitual slips, at
the CCEP meeting. However, it is difficult to obtain fine images of small
slips, particularly in inland, even though inland GPS stations are arranged
closely. This is attributed to high seismicity rather than GPS observation
errors. Because strong earthquakes occur frequently, various effects of
slow slips in GPS records are mixed up with such stronger changes. Hence,
development of statistical models and methods to separate such signals is
crucial. To estimate slow slips more precisely, combined modeling and
analysis of seismicity and geodetic anomalies will be useful. Analyzing
both seismicity and transient geodetic movements in a number of areas and
locating the area of aseismic slip is very important for increasing the
probability gain of a large earthquake.

\subsection{Considering the Scenario of an Earthquake from Aseismic
Slip}\label{sec6.2}

Observed anomalies of crustal movement and seismicity assume fault
mechanisms and locations of slip precursors for prediction probability;
therefore, their uncertainty must be estimated. In addition, probabilities
of considered scenarios must be estimated. Such tasks are difficult. A
possible method is to consider the logic tree of various scenarios
regarding destruction of the fault system by attaching appropriate
subjective or objective probabilities to tree components, as was performed
for long-term predictions in California and Japan. Hence, such a scenario
ensemble gives a forecast probability. Similarly, medium- and short-term
prediction logic trees of various scenarios must be considered.

At the CCEP meeting on April 6, 2005, I reported relative quiescence of
aftershocks of the Fukuoka--Oki earthquake (Ogata, \citeyear{OGA05d}). In addition, I
examined potential slow slip areas on nearby active faults that may have
created a stress shadow (region of decreased CFS) to cause relative
quiescence in the aftershock sequence. Among these areas, the Kego fault,
traversing Fukuoka City, had a large positive $\Delta$CFS because of the
main shock rupture, which showed evidence of possible slow-slip induction.
Furthermore, the seismogenic zone along the Kego fault had already
activated before the Fukuoka--Oki earthquake occurred (\cite{OGA10N1}).
Therefore, because a slow-slip scenario on this fault was possible, I
examined the pattern causing stress variation in the aftershock area.
However, no stress shadow in the aftershock area was found. Therefore, I
determined that probability of slow slip on the Kego fault was quite low. I
also examined whether other possible slow slips in neighboring active
faults could create a stress shadow that covered the aftershock region.
However, no large earthquake has occurred in those faults thus far.

Approximately one month later, however, the largest aftershock occurred at
the southeast end of the aftershock zone. Post-mortem examination based on
information of the fault mechanism of this aftershock and detailed
aftershock data revealed a detailed scenario. This means that by assuming a
slow slip into the gap between the fault of the largest aftershock and main
shock, relative quiescence of activity in the deeper part of the aftershock
zone can be clearly explained (\cite{OGA06}). Moreover, the slip can explain
relative quiescence in the induced swarm activity that occurred away from
the aftershock area (\cite{OGA06}).

This setting as a prediction of future scenarios is much more vague and
difficult to explain, even if it includes an ex-post scenario. Moreover,
the time of occurrence must be predicted in addition to location, which is
more difficult. Occurrence of slow slip does not always indicate a
proximate precursor of fault corruption. Nevertheless, it is desired to
keep observing GPS data to form scenarios of forthcoming large earthquake.
For example, using Bayesian inversion by using GPS records, \citet{HASetal09}
estimated\vadjust{\goodbreak} the locked zones on the plate boundary, where the next
great earthquakes are expected.

\section{Conclusions}\label{sec7}

To predict the future of a complex and diverse earthquake generation
process, probability forecasting cannot be avoided. The likelihood
(log-likelihood) is rational to measure the performance of the prediction.
To provide a standard stochastic prediction of seismic activity in long
term and short term, it is necessary to construct proper point process
models and revise those that conform to each region. By the appearance of
the anomaly, we need to evaluate the probability that it will be a
precursor to a large earthquake. Namely, we need to forecast that the
probability in a space--time zone will increase to an extent, relative to
those of the reference probability. For this, we make use of a point
process model for the causality relationship.

It is desired to search any anomaly phenomena that enhance the probability
gains. Having such anomalies, application of the multiple element
prediction formula increases a precursory probability. A comprehensive
physical study between precursory phenomena and earthquake mechanisms is
essential for composing useful point process models. These elements must be
incorporated to achieve predicted probability exceeding predictions of
typical statistical models.

Furthermore, to determine urgency and uncertainty of major earthquakes
against abnormal phenomena, numerous research examples must be accumulated.
On the basis of these examples, possible prediction scenarios must be
presented. Furthermore, to adapt well to diversity of earthquake
generation, it is useful to adopt Bayesian predictions (\cite{Aka80N2}; \cite{NOMetal11}) and consider region-specific models.

My experiences thus far confirm that the method of statistical science is
essential to elucidate movement leading to prediction of a complex system.
There is a need for development of a forecasting model that reflects
diversity of the vast amount of information on seismicity and various
covariate data. I believe that these will be developed by inventing an
appropriate hierarchical Bayesian model. Space--time models for seismicity
have become increasingly complicated (Ogata, \citeyear{OGA98}, \citeyear{OGA04N1}, \citeyear{OGA11N1}; \cite{OgaKatTan03}; \cite*{OGAZHU06}).

A similar evolution is required for statistical models of geodetic GPS
data.

\section*{Acknowledgments}

I am grateful to the Japan Meteorological Agency (JMA), the
National Research Institute for Earth Science and Disaster Prevention, and
universities for the providing hypocenter data. I am also grateful to the
anonymous referee, Associate Editor and the Editor for their careful
reviews and suggestions, which led to a significant revision of the present
manuscript.
This work was supported by JSPS KAKENHI Grant Number 23240039, and by the Aihara
Innovative Mathematical Modelling Project, the ``Funding Program for
World-Leading Innovative R\&D on Science and Technology (FIRST Program),''
initiated by the Council for Science and Technology Policy.


%
%
%
%
%
%
%
%



\end{document}